\documentclass[%
prb,
twocolumn,
superscriptaddress,
 amsmath,amssymb,
 aps,
]{revtex4-1}

\usepackage{graphicx}
\usepackage{dcolumn}
\usepackage{bm}

\begin{document}

\title{Environment-Assisted Quantum Transport through Single-Molecule Junctions}

\author{Jakub K. Sowa}
\email{jakub.sowa@materials.ox.ac.uk}
\author{Jan A. Mol}
\author{G. Andrew D. Briggs}
\affiliation{%
 Department of Materials, University of Oxford, Parks Road, Oxford OX1 3PH, United Kingdom
}%


\author{Erik M. Gauger}
\affiliation{SUPA, Institute of Photonics and Quantum Sciences, Heriot-Watt University, EH14 4AS, United Kingdom}



\date{\today}
\begin{abstract}
Single-molecule electronics has been envisioned as the ultimate goal in the miniaturisation of electronic circuits. While the aim of incorporating single-molecule junctions into modern technology still proves elusive, recent developments in this field have begun to enable experimental investigation of fundamental concepts within the area of chemical physics. One such phenomenon is the concept of Environment-Assisted Quantum Transport which has emerged from the investigation of exciton transport in photosynthetic complexes. Here, we study charge transport through a two-site molecular junction coupled to a vibrational environment. We demonstrate that vibrational interactions can significantly enhance the current through specific molecular orbitals. Our study offers a clear pathway towards finding and identifying environment-assisted transport phenomena in charge transport settings.
\end{abstract}

\maketitle

\section{Introduction}
Energy transport in photosynthetic complexes critically depends on the coupling between the electronic and environmental (vibrational) degrees of freedom~\cite{ishizaki2009theoretical,ishizaki2010quantum}. It has been suggested that these interactions can significantly enhance the efficiency of exciton transport \textit{in vivo}. 
This phenomenon, referred to as Environment-Assisted Quantum Transport (ENAQT)~\cite{mohseni2008environment,plenio2008dephasing,rebentrost2009environment,chin2010noise}, has attracted a great deal of attention in the nascent field of quantum biology. 
Vibrational interactions can augment the energy transport in at least two ways: First, by assisting transitions across energy gaps, and second, by inhibiting destructive interference~\cite{caruso2009highly}. For structures with high degree of symmetry, they could also boost transport performance via the mechanism of momentum rejuvenation~\cite{li2015momentum}. 

In principle, we expect the same mechanisms to play an analogous role in charge transport through molecular systems~\cite{ballmann2012experimental}. Here, single-molecule junctions (SMJs) offer an interesting alternative to the typical ensemble spectroscopic measurements in charge and energy transfer studies~\cite{engel2007evidence,panitchayangkoon2011direct,renaud2016deep}. In these highly controllable systems, the `transport efficiency' can be quantified directly as the steady-state current passing through the system. SMJs are nowadays routinely fabricated and studied using carefully tuned bias and back-gate potentials over an impressive temperature range, providing an excellent platform for proof-of-principle experiments.

In contrast to an extensive body of literature on off-resonant transport through molecular wires (via the co-tunnelling or vibrationally-assisted mechanism)~\cite{segal2000electron,kilgour2015charge,kim2017controlling}, here we shall focus on the {\it resonant regime} where (otherwise unitary) dynamics within the molecular system is modulated by environmental interactions. As we will discuss, this fundamentally different physical regime offers new perspectives for studying the interplay of unitary and dissipative dynamics. 

\section{Theoretical model}
\begin{figure}[ht]
\centering
  \includegraphics{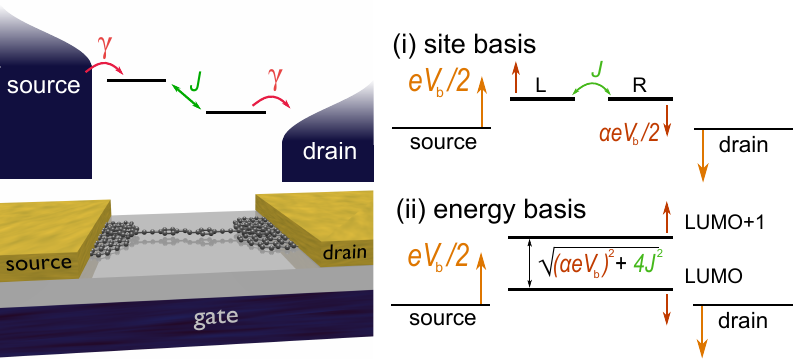}

  \caption{Left: schematic depiction of our system. Right: level structure in site and energy basis; arrows indicate energy shifts under an applied bias voltage.}
  \label{schemeI}
\end{figure}

We consider a two-site molecular system where each of the sites (L and R) couples to an independent phonon bath and either the source or the drain electrode, schematically depicted in Fig.~\ref{schemeI}. The two-site character of the molecule can be most easily achieved by breaking the conjugation within the system (introducing regions of low $\pi$-electron density). Several structures of this type have recently been investigated in the transport setting~\cite{guedon2012observation,perrin2016gate,perrin2014large,taherinia2016charge,koole2016spin,kaliginedi2012correlations}; the model used here is inspired by the experimental studies of Perrin \textit{et al.}~\cite{perrin2014large,perrin2016gate}. 

We now proceed to describe the Hamiltonian governing our system (with $\hbar=1$ and $e=1$ throughout).
The molecular tight-binding Hamiltonian is given by:
\begin{equation}
H_{\mathrm{M}} = \sum_{j=\mathrm{L},\mathrm{R}} \varepsilon_j\: a^\dagger_j a_j + J( a^\dagger_{\mathrm{L}} a_{\mathrm{R}} + \mathrm{H.c.}) ~,
\end{equation}
with H.c.~denoting the Hermitian conjugate, while $a^\dagger_j$ ($a_j$) is the usual raising (lowering) operator which creates (annihilates) an electron on site $j$ with energy $\varepsilon_j$ (with respect to the Fermi energy of the leads), and $J$ is the strength of the tunnel coupling between the neighboring sites. As demonstrated experimentally, the site energies are sensitive to an applied bias voltage~\cite{perrin2014large,perrin2016gate}. We will focus on a symmetric molecular structure in which the sites capacitively couple to the source or drain electrode as follows: $\varepsilon_{\mathrm{L}} = \varepsilon_0 + \alpha V_{\mathrm{b}}/2$ and  $\varepsilon_{\mathrm{R}} = \varepsilon_0 - \alpha V_{\mathrm{b}}/2$ for an applied bias of $V_{\mathrm{b}}$, so that $\alpha V_{\mathrm{b}}$ is the voltage drop within the molecule. 
It should be recognised that $\alpha$ depends on the microscopic details of the junction, and will vary between different single-molecule devices.
However, as values between 0.48 and 0.74 have been reported in experimental and \textit{ab initio} studies~\cite{perrin2014large,perrin2016gate,perrin2015single}, we shall henceforth adopt $\alpha = 0.6$ unless stated otherwise.

As the site-orbitals are not eigenstates of the molecular Hamiltonian, they hybridise into two linear (bonding and antibonding) combinations -- or molecular orbitals (MO) -- at energies:
$\varepsilon_{\pm} =  \varepsilon_0 \mp \sqrt{(\alpha V_b)^2 + 4 J^2}/2 ~.$

We treat the environment as bosonic baths of (thermalised) vibrational modes of frequencies $\omega_{q_j}$. Given raising (lowering) operators $b^\dagger_{q_j}$ ($b_{q_j}$), the bath Hamiltonian reads
\begin{equation}
H_{\mathrm{B}} = \sum_{j=\mathrm{L},\mathrm{R}}\sum_{q_j} \omega_{q_j} b^\dagger_{q_j} b_{q_j}~,
\end{equation}
and the modes couple to the electronic degrees of freedom within the molecular wire with strength $g_{q_j}$ via
\begin{equation}
H_{\mathrm{C}} = \sum_{j=\mathrm{L},\mathrm{R}}\sum_{q_j} g_{q_j} a^\dagger_j a_j(b^\dagger_{q_j} + b_{q_j})~.
\end{equation}
We shall make use of the usual definition of the spectral density (SD) to characterise the environmental coupling:
$\mathcal{J}_j(\omega) = \sum_{q_j} \;\lvert g_{q_j}\rvert^2 \;\delta (\omega - \omega_{q_j})~.$
Note that we have assumed that each site interacts with its own independent phononic environment. Considering a (partially) shared environment is also possible but may result in additional phenomena~\cite{sowa2017vibrational, nazir2009correlation} which would, however, detract from the core message of this study.
Finally, the fermionic reservoirs (leads) and their couplings to the molecular sites are governed by the Hamiltonians:
\begin{align}
H_{\mathrm{R}} &= \sum_{j=\mathrm{L},\mathrm{R}}\sum_{k_j}\epsilon_{k_j} c_{k_j}^\dagger c_{k_j} ~,\\
H_{\mathrm{V}} &= \sum_{k_{\mathrm{L}}, k_{\mathrm{R}}} V_{k_{\mathrm{L}}} c_{k_{\mathrm{L}}}^\dagger a_{\mathrm{L}} + V_{k_{\mathrm{R}}} c_{k_{\mathrm{R}}}^\dagger a_{\mathrm{R}} + \mathrm{H.c.} ~,
\end{align}
where $c_{k_j}^\dagger$ ($c_{k_j}$) creates (annihilates) an electron in the lead level $k_{j}$. 
The overall Hamiltonian is then given by:
$H = H_{\mathrm{M}} + H_{\mathrm{B}} + H_{\mathrm{C}} + H_{\mathrm{V}} + H_{\mathrm{R}}~.$

The details of our theoretical calculations are comprehensively discussed in the ESI$^\dag$. Working in the limit of strong Coulomb blockade, we treat the molecule-lead interactions perturbatively within the Born-Markov \cite{breuer2002theory} and the wide-band approximation ($V_{k_j} = V_j = \mathrm{const.}$ ). The interactions with the vibrational environment will be accounted for using three different theoretical approaches briefly described below. 
All of them yield a quantum master equation (QME) for the evolution of the reduced density matrix $\rho(t)$ of the form:
\begin{equation}\label{QME}
\dfrac{\mathrm{d}\rho(t)}{\mathrm{d}t} =-\mathrm{i}[H_{\mathrm{S}},\rho(t)] + \mathcal{L}_{\mathrm{leads}} \rho(t) + \mathcal{L}_{\mathrm{ph}} \rho(t) ~,
\end{equation}
where $H_{\mathrm{S}}$ is the system Hamiltonian, and $\mathcal{L}_{\mathrm{leads}}$ and $\mathcal{L}_{\mathrm{ph}}$ are superoperators describing the coupling between the molecular electronic levels and respective environments. Eq.~\eqref{QME} is solved in the steady-state limit, $\mathrm{d}\rho(t)/\mathrm{d} t =0$, and the average electric current is obtained as an expectation value of the current superoperator ($\mathcal{I}$) in the steady state, $I = \mathrm{Tr}[\mathcal{I}\rho_{\mathrm{st}}]$~\cite{flindt2005full} (also see ESI$^\dag$).
Besides average current, the Fano factor is an important (and measurable \cite{karimi2016shot}) observable of interest, quantifying the deviation of zero-frequency current noise, $S(0)$, from Poissonian noise: $F = S(0)/2eI$~\cite{blanter2000shot}.


\section{Results and Discussion}
Throughout this work we set $\varepsilon_0 = 27$ meV, $J=-24$ meV, and $\gamma_{\mathrm{L}}=\gamma_{\mathrm{R}} =1$ meV (where $\gamma_j = 2\pi \varrho_j \lvert V_j\rvert^2$, $\varrho_j$ being the constant density of states in lead $j$). The bonding and antibonding MOs both lie above the Fermi energy and will henceforth be referred to as the LUMO and LUMO+1 respectively\footnote[3]{Here, we consider a pair of MOs lying above the Fermi energy of the leads. At zero bias, they are hence the two lowest unoccupied MOs of the molecular system even though they do not necessarily have to correspond to the vacuum LUMO and LUMO+1 levels.}. 
The choice of the site energies above is not of critical importance, particularly since many currently available experimental techniques allow electrostatic control of the molecular energy levels through a gate electrode~\cite{mol2015graphene,gehring2017distinguishing,burzuri2016sequential,koole2016spin}. 

\begin{figure}[ht]
\centering
  \includegraphics{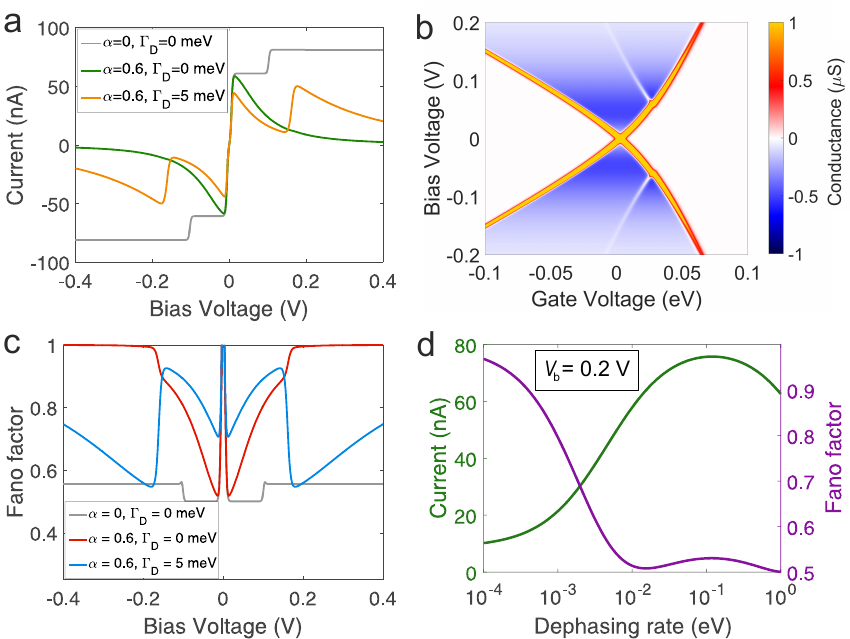}
  \caption{(a) $IV$ characteristics; (b) Conductance map for $\alpha=0.6$ in the absence of environmental coupling; (c) Fano factors as a function of $V_\mathrm{b}$; (d) Current values, and Fano factor at $V_{\mathrm{b}} = 0.2$ eV as a function of $\Gamma_\mathrm{D}$. The temperature in the leads: $T = 10$ K.}
  \label{fig2}
\end{figure}

We begin by considering transport in the absence of any vibrational coupling. The $IV$ characteristics for this case are shown in Fig.~\ref{fig2}(a) for different values of $\alpha$. The presence of two current steps (for $\alpha=0$) reveals the existence of two transport channels: one for each of the MOs included in our model.The first plateau spans bias voltage range at which only the LUMO level is located within the bias window (its width corresponds thus to twice the energy gap between LUMO and LUMO+1). As discussed above, for non-zero $\alpha$ 
an applied bias energetically detunes the two site orbitals. As a result, the efficiency of transport through each of the channels (as quantified by the electric current) decreases with increasing bias leading to Negative Differential Conductance (NDC)~\cite{xu2015negative}. This effect has been observed experimentally and discussed by Perrin \textit{et al.}~\cite{perrin2014large}. Note that the capacitive coupling shifts the position of the LUMO/LUMO+1 levels and hence also the position of the steps in the $IV$ characteristics. A calculated conductance map (differential conductance as a function of the bias and gate voltage) for the studied molecular system is shown in Fig.~\ref{fig2}(b). Therein, for simplicity, we have taken $\varepsilon_{\mathrm{L/R}} = \varepsilon_0 \pm \alpha V_{\mathrm{b}}/2 - V_{\mathrm{g}}$, where $V_{\mathrm{g}}$ is the back-gate potential. Similarly to Fig.~\ref{fig2}(a), it demonstrates the existence of two transport channels showing pronounced NDC features. Importantly, the edges of the Coulomb diamonds and the lines corresponding to the higher excited state are curved (in the opposite directions for the LUMO and LUMO+1 levels) due to the capacitive coupling of the sites to the source and drain electrodes. Note that while the differential conductance is symmetric with respect to the bias voltage it is not symmetric with respect to the applied gate potential.

We now consider the role of vibrational coupling, initially by employing pure dephasing as the simplest, albeit widely used, way of capturing environmental interactions~\cite{contreras2014dephasing,kilgour2015tunneling,plenio2008dephasing,rebentrost2009environment}. In this approach, the role of environmental coupling is reduced to an exponential decay of the off-diagonal elements of the density matrix in the site basis. 
The result of introducing phenomenological dephasing (at rate $\Gamma_{\mathrm{D}}$) is shown in Fig.~\ref{fig2}(a). Firstly, the electric current through the LUMO orbital is significantly suppressed. Introducing pure dephasing results in a population transfer between the LUMO and LUMO+1 orbitals. Since at low bias the LUMO+1 level lies outside of the bias window, charge density transferred into this level drains into both leads. This leads to a decrease in net current flowing through the system.    
Secondly, the current flowing through both MOs (at $V_{\mathrm{b}}$ above $\sim 0.17$ V when $V_{\mathrm{b}}/2 = \varepsilon_-$) significantly increases with $\Gamma_{\mathrm{D}}$. This is an example of an environment-assisted phenomena which is the focus of this work. At high bias voltage there exists a large energy gap between the two site orbitals rendering the unitary $\lvert \mathrm{L}\rangle \leftrightarrow \lvert \mathrm{R}\rangle$ transition highly inefficient. The presence of dephasing projects population onto specific sites and thus helps to overcome energy steps. 
The values of current at $V_{\mathrm{b}}=0.2$ V as a function of the dephasing rate are shown in Fig. \ref{fig2}(d).
The efficiency of transport through the molecule increases with $\Gamma_{\mathrm{D}}$ as long as it does not considerably exceed the system's characteristic frequency (the energy difference between the MOs, here $\sim 0.13$ eV), beyond that point $I$ decreases with the dephasing rate eventually reaching the Quantum Zeno limit~\cite{rebentrost2009environment}.
The conditions necessary to observe environment-assisted transport effects in our model system can be summarised as follows: firstly, the dephasing rate must be at least comparable to the molecule-lead coupling strengths, and secondly, the bias-induced energy gap between the sites ($\alpha V_{\mathrm{b}}$) should be greater than the strength of the inter-site coupling $J$. We expect this regime to apply to a wide range of molecular multi-site structures especially if studied in an appropriate device geometry.

Fig.~\ref{fig2}(c) shows the Fano factor as a function of the bias voltage. For $\alpha=0$ and at low bias the Fano factor takes the value of roughly $1/2$
(characteristic of a single transport channel), whereas at high bias it approaches $F \approx 5/9$
(the expected infinite-bias value for a strongly coupled resonant double quantum dot)~\cite{kiesslich2007noise,elattari2002shot,blanter2000shot}. Energetic detuning of the sites results in an increase of the Fano factor, tending towards unity in the high bias limit. Likewise, for transport occurring solely through the LUMO, $F$ increases in the presence of pure dephasing which is again attributable to environment-induced population transfer between the molecular levels.
By contrast, pure dephasing decreases $F$ at high bias in accordance with studies on resonant tunnelling diodes~\cite{kiesslich2007noise,stones2017non}, see Fig.~\ref{fig2}(d).

While the pure dephasing approach possesses the appeal of simplicity, its phenomenological nature and  infinite-temperature character can lead to (unphysical) behaviour which differs substantially from microscopically founded alternatives (see the ESI$^\dag$ for a full discussion of the limitations of pure dephasing in the context of our model). 
Treating environmental effects more rigorously requires making certain assumptions about the nature of the vibrational coupling: for simplicity, we assume that the electronic degrees of freedom (DoFs) interact with an unstructured environment and use
a superohmic SD with exponential cut-off:
\begin{equation}
\mathcal{J}(\omega) = \dfrac{\lambda}{2} \; \dfrac{\omega^3}{\omega_{\mathrm{c}}^3} \; e^{-\omega/\omega_{\mathrm{c}}}~.
\end{equation}
Here, $\lambda$ is the reorganisation energy and $\omega_{\mathrm{c}}$ is the cut-off frequency. This parameterisation conveniently separates the reorganisation energy from $\omega_\mathrm{c}$, however, note that dissipative Redfield rates depend on both $\lambda$ and $\omega_\mathrm{c}$ even for low-frequency transitions (unlike, e.g., for the spectral densities of Refs. \cite{ramsay2010phonon,gauger2010heat}). 
Electronic DoFs in molecular systems are generally coupled to intra-molecular vibrational modes~\cite{galperin2007molecular,galperin2006resonant,lau2015redox,burzuri2016sequential,paulsson2006inelastic,okabayashi2010inelastic,nazin2005tunneling,pradhan2005vibronic} as well as a wider environment (for instance a solvent or a substrate on which the molecule is deposited). The resulting damped molecular modes could be accounted for using the approach of `tiered environments'~\cite{fruchtman2015quantum}, or alternatively, modelled through an effective continuous SD~\cite{roden2012accounting}. We here adopt the latter approach and note that
structured spectral densities (for instance motivated by specific molecular systems) may be required for quantitative agreement with experimental data, but this is a trivial extension of our approach. 

In the remainder of this work, we will use two different theoretical approaches, which we henceforth refer to as the Redfield, and the Polaron methods. Both of them utilise non-secular Born-Markov approximations~\cite{breuer2002theory,nazir2016modelling} which in the case of the Redfield technique is applied directly to the vibrational coupling ($H_{\mathrm{C}}$).
By contrast, the Polaron method relies on an initial Lang-Firsov transformation to eliminate electron-phonon coupling terms from the Hamiltonian -- at the expense of introducing displacement operators to the $H_{\mathrm{V}}$ and $H_{\mathrm{M}}$ terms~\cite{nazir2016modelling,lang1963kinetic,jang2008theory}. The Polaron approach is not only more accurate for stronger system environment coupling (i.e.~larger $\lambda$) but also captures vibrational effects at the lead-molecule interfaces.
Let us note here that the Polaron technique is, in fact, also capable of describing purely classical dynamics: for strong enough environmental coupling and at high temperature, transport is solely mediated by dissipative terms in the Polaron QME (reducing to Marcus theory~\cite{marcus1985electron} under certain assumptions -- see ESI$^\dag$).

\begin{figure}[ht]
\centering
  \includegraphics{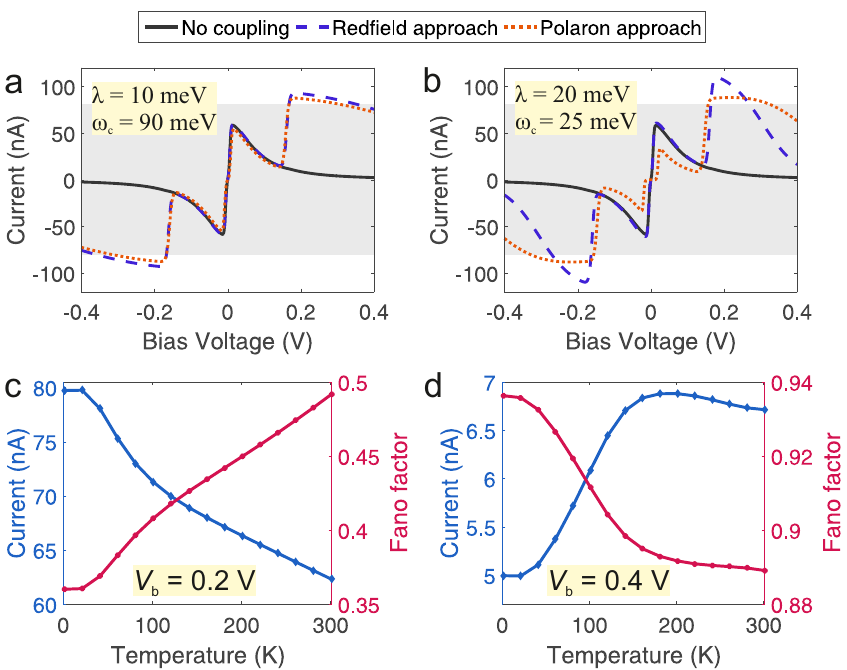}
  \caption{(a,b) $IV$ characteristics in the absence of environmental coupling, and presence of coupling to superohmic phonon baths for different values of the $\lambda$ and $\omega_\mathrm{c}$ at $T=10$ K. (c,d) Values of steady-state current and the Fano factor as a function of temperature in the case of coupling to phonon baths with $\lambda = 10$ meV, $\omega_{\mathrm{c}} = 15$ meV obtained using the Polaron method.}
  \label{fig3}
\end{figure}

The $IV$ characteristics calculated using these microscopic approaches are presented in Fig.~\ref{fig3}.
As shown therein, the two methods yield closely coinciding results for relatively weak coupling. The small differences between the two approaches in this parameter regime stem predominantly from phonon broadening (effects on molecule-lead interfaces, see ESI$^\dag$) that are not captured by the Redfield method. For this case, we also note a significant qualitative resemblance with the results obtained using the pure dephasing model.
By contrast, in the strong coupling regime the two methods deliver markedly different results, see Fig.~\ref{fig3}(b). We attribute this partially to considerable phonon broadening, and partially to a failure of the perturbative Redfield technique in the stronger coupling regime.

Employing microscopic methods (as opposed to the phenomenological pure dephasing) also allows us to properly account for the temperature dependence of the observables of interest. Fig.~\ref{fig3}(c) shows the values of current and the Fano factor as a function of temperature for different values of the bias voltage calculated using the Polaron method. At lower bias, $V_\mathrm{b}=0.2$ V, the current decreases with temperature as it is dominated by effects occurring at the molecule-lead interfaces: phonon and Fermi broadening~\cite{poot2006temperature}. 
The opposite  trend is present at high bias $V_\mathrm{b}=0.4$ V, Fig.~\ref{fig2}(d), where transport is largely insensitive to interfacial effects, and an increase in temperature augments intramolecular electron hopping. (Even there, however, Fermi and phonon broadening will eventually lead to a decrease in current at high enough $T$.) In both cases the Fano factor displays a temperature dependence that is anti-correlated with that of the steady-state current. We conclude that in realistic systems the response of the current as well as Fano factor to increasing temperature will \textit{qualitatively} depend on the bias voltage at which it was measured, even in the resonant tunnelling regime.

We return to Fig.~\ref{fig3}(a) and (b) and focus on the magnitude of the current maxima: surprisingly, environmental coupling paired with non-zero $\alpha$ (i.e.~detuned sites) lets the LUMO+1 current peaks exceed the large-bias plateau of the resonant $\alpha=0$ case (in the absence of environmental interactions). This bound is indicated by the grey background shading in Figs.~\ref{fig3}(a,b).
\begin{figure}[ht]
\centering
  \includegraphics{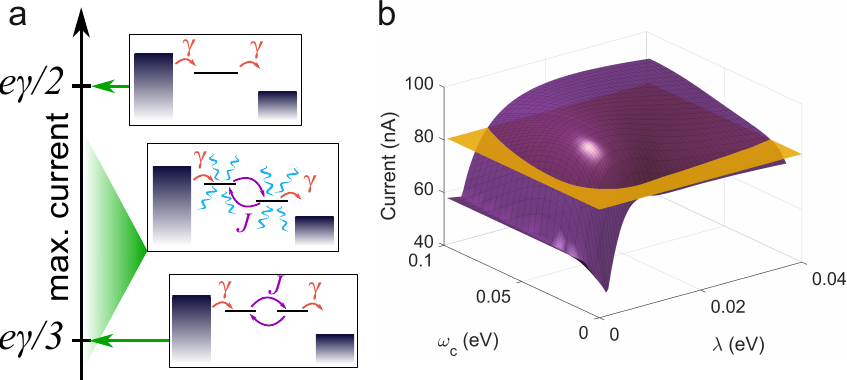}
  \caption{(a) Schematic illustration, showing the available current maxima for different scenarios with $J \gg \gamma$ where $\gamma=\gamma_{\mathrm{L}} =\gamma_{\mathrm{R}}$. (b) Overall maximum value of the current as a function of  $\lambda$ and  $\omega_{\mathrm{c}}$ at $T=10$ K (Polaron method). The orange plane corresponds to maximum current for $\alpha=0$ and $\lambda=0$. }
  \label{fig4}
\end{figure} 
A common interpretation of ENAQT is that it regains (some of) the efficiency lost due to the energetic network disorder~\cite{rebentrost2009environment}. Clearly, here a combination of detuning and environmental coupling unlocks a maximal current which {\it exceeds} what is available from the archetypal idealised quantum channel~\cite{caruso2014universal} (i.e.~a noiseless degenerate chain for end-to-end transport).
In the limit of $J \gg \gamma$, i.e.~for comparatively weak
molecule-lead coupling, a quasi-equilibrium is established following an electron jump from the source onto the molecule, before the electron exits into the drain. In the degenerate case the additional charge density populates equally both of the sites. On the other hand, energetic detuning together with environmental coupling favours localisation of the charge density on the site with lower energy thus enhancing the steady-state transport efficiency.
However, we note that the current through a two-site system can never surpass that of an equivalent single-site system, Fig.~\ref{fig4}(a)\footnote[4]{For a single-site system, if $\gamma_\mathrm{L} = \gamma_\mathrm{R}=\gamma$, the steady-state population of the additional charge density on the molecule is $\frac{1}{2}$. This gives an average current of $e\gamma/2$. For a two-site system, where $\gamma\ll J$, the steady-state population on each of the sites is $\frac{1}{3}$ which yields an average current of $e\gamma/3$.}.

This phenomenon is depicted more clearly in Fig.~\ref{fig4}(b) which shows the overall maximum value of current as a function of the reorganisation energy and the cut-off frequency of the phonon baths as compared to the value for $\alpha=0$ and $\lambda=0$ (orange plane). We note that qualitatively similar behaviour occurs for weaker coupling, and that the magnitude of the effect depends on the environmental coupling strength [\textit{c.f.} Fig.~\ref{fig3}(a)]\footnote[5]{The agreement between the Redfield and the Polaron approach, which are justified in complementary parameter regimes~\cite{mccutcheon2011general}, underlines that this effect is not just an artefact arising from the approximations involved in the derivation of a particular model.}.

\section{Conclusions}
We have applied the paradigm of environment-assisted quantum transport in a charge transport setting, using the example of a two-site molecular junction coupled to a phononic environment. We have shown that environmental coupling significantly increases the electric current flowing through simple molecular systems, by assisting charge propagation across an energy gap within the molecule, typically accompanied by a decrease in the Fano factor. 
While the effect of environmental interactions increases with temperature, the temperature-dependence of current and the Fano factor varies greatly with the bias voltage at which it is measured.
Finally, we predict that a combination of energetic detuning and environmental coupling unlocks steady-state currents beyond what is available for idealised purely resonant transfer. This observation goes beyond the typical interpretation of ENAQT of mitigating the effects of energetic disorder.

Having employed three different methods to capture the effect of the environment, we have shown that the key phenomena we discuss are inherently robust. This suggests they should persist over a wide range of systems and parameter regimes, provided the strength of the environmental interactions is at least comparable with electronic coupling strengths. This offers exciting opportunities for experimental investigation of the intriguing physics occurring at the quantum--classical interface.

\begin{acknowledgments}
The authors thank Colin Lambert for useful discussions.
J.K.S. also thanks the Clarendon Fund and EPSRC for financial support. E.M.G. acknowledges funding from the Royal Society of Edinburgh and the Scottish Government, J.A.M. acknowledges funding from the Royal Academy of Engineering. This work was supported by the EPSRC QuEEN Programme Grant (EP/N017188/1) and the John 
Templeton Foundation.
\end{acknowledgments}

\end{document}